\documentclass[prl,aps,twocolumn,showpacs]{revtex4}

\usepackage{epsfig}

\begin{document}

\title{Is La$_{1.85}$Y$_{0.15}$CuO$_{4}$ an oxygen-doped cuprate superconductor?} 

\author{W. Yu$^{1}$}
  \email{weiqiang@squid.umd.edu}
\author{B. Liang$^{1}$}
\author{P. Li$^{1}$}
\author{S. Fujino$^{1,2}$}
\author{T. Murakami$^{1,2}$}
\author{I. Takeuchi$^{1,2}$}
\author{R. L. Greene$^{1}$}
\affiliation{$^1$Center for Superconductivity Research, Department of Physics, \\
$^2$Department of Material Science and Engineering,\\ 
University of Maryland, College Park, MD 20742}

\date{\today}

\begin{abstract}

We report resistivity, Hall effect, Nernst effect, and magnetoresistance measurements on T'-phase La$_{1.85}$Y$_{0.15}$CuO$_{4}$ (LYCO) films prepared %%@
by pulsed laser deposition under different oxygen conditions. Our results show that superconductivity in LYCO originates from an oxygen-doped %%@
Mott-like insulator and not from a weakly correlated, half-filled band metal as proposed previously. 
  
\end{abstract}

\maketitle

The origin of the high-temperature superconductivity in doped copper oxides is one of the major unresolved problems in cuprates. The parent (undoped) %%@
compounds, e.g., La$_2$CuO$_4$, are predicted to be half-filled band metals by simple non-interacting electron band models. But experimentally all the %%@
copper oxides parent compounds prepared to date are known to be antiferromagnetic (AFM) insulators. Upon doping, superconductivity is achieved with a %%@
suppression of antiferromagnetism. This has led to the general belief that electron-electron correlations play a very important role in the normal %%@
state and the superconducting properties of this class of copper oxides. The simplest model used to explain the spin=1/2 antiferromagnetism on copper %%@
sites is the Mott-Hubbard model \cite{Anderson}. Most of the theories attempting to explain the origin of superconductivity in the copper oxides have %%@
started from some variation of this basic model for strongly correlated systems.   

Recently, a very surprising result was reported, which claimed that the parent, undoped copper oxide system T'-structure (La, RE)$_{2}$CuO$_{4}$ %%@
(LRCO) is a band metal and not a Mott-like insulator \cite{Naito_SSC, Noda_PC_426_220, Tsukada_PC_426_459}. Small rare earth ions RE$^{3+}$, such as %%@
Y$^{3+}$, Tb$^{3+}$, {\it etc.}, were partially substituted for La$^{3+}$ using a MBE thin film technique. This keeps the total charge the same as in %%@
the T'-phase La$_2$CuO$_4$, and implies that no doping occurs. These materials are metastable and cannot be made by conventional bulk growth methods. %%@
A very thorough study \cite{Naito_SSC, Noda_PC_426_220, Tsukada_PC_426_459} of the RE doping, preparation conditions, and the properties of these LRCO %%@
materials was made. At some RE dopings, superconductivity with Tc's as high as 23K was found! This was explained as arising from a weakly correlated, %%@
half-filled, band metal state. Since this work questions one of the most fundamental assumptions about the possible origin of high-temperature %%@
superconductivity in the cuprates, i.e., the Mott-Hubbard parent state, it is quite important to confirm these new experimental results and their %%@
interpretations. 

In this paper, we show that the doping of a Mott-like insulator is the origin of the physical properties in the LRCO system. Our resistivity, Hall %%@
effect, Nernst effect and magnetoresistance data suggest that oxygen reduction of La$_{1.85}$Y$_{0.15}$CuO$_{4}$ (LYCO) films is responsible for the %%@
metallic and superconducting properties. Since oxygen reduction is equivalent to cerium doping in the T'-phase Ln$_{2-x}$Ce$_{x}$CuO$_{4}$ (LnCCO), we %%@
believe LYCO is an electron-doped superconductor which evolves from a Mott-Hubbard AFM insulating state. Evidence for a spin-density-wave (SDW) or AFM %%@
state in the as-grown LYCO films, and a Fermi surface evolution from an electron-like to a two-band system with increasing oxygen reduction were %%@
observed. This is the same behavior found in cerium-doped LnCCO materials. Therefore, our data are incompatible with the scenario of a half-filled %%@
band metal.

\begin{figure}
\includegraphics[width=7cm, height=7cm]{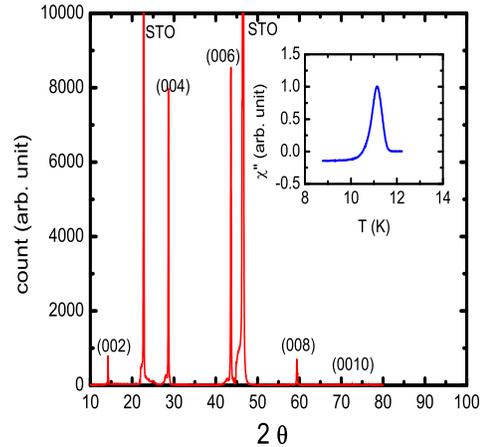}
\caption{\label{x-ray} The X-ray diffraction pattern of a c-axis oriented La$_{1.85}$Y$_{0.15}$CuO$_{4- \delta}$ film (sample A7). Inset: The %%@
imaginary ac susceptibility $\chi ''$ vs. temperature.}
\end{figure}

Our c-axis oriented LYCO films were grown by the pulsed laser deposition technique, using a stoichiometric La$_{1.85}$Y$_{0.15}$CuO$_{4}$ ceramic as a %%@
target. The films were deposited first on STO (SrTiO$_{3}$) or KTO (KTaO$_3$) substrates at 700$^\circ$C under 230 mTorr N$_2$O pressure, then %%@
annealed at 620$^\circ$C {\it in situ} under vacuum of $10^{-6}$ Torr, and finally cooled in the vacuum down to room temperature. Vacuum annealing of %%@
the T'-structure cuprates is known to remove oxygen from the sample (i.e., oxygen reduction). Under these conditions, a c-axis oriented T'-phase LYCO %%@
is well established. Typical film thickness is about $2500\AA$ and  $T_C$ depends on the annealing time and pressure (see Table ~\ref{sample}). The %%@
X-ray diffraction pattern and ac susceptibility of a superconducting film (sample A7) are shown in Fig.~\ref{x-ray}, indicating the T'-phase %%@
($c\approx 12.4 \AA$) and bulk superconductivity. To our knowledge, this is the first time the standard PLD method has been used without a buffer %%@
layer to achieve La-based T'-phase superconducting films. The transport properties are similar for films made with an STO substrate or a KTO substrate %%@
although the annealing time is slightly different. The resistivity of our films is comparable to those made by MBE \cite{Tsukada_PC_426_459}. Our %%@
ab-plane resistivity, Hall-effect, and magnetoresistance measurements were performed on Hall-bar patterned films in a Quantum Design PPMS. Nernst %%@
effect measurements were conducted in the PPMS with a home-built Nernst setup.

\begin{table}
 \caption{Comparison of the bulk superconducting transition temperature T$_C$ (measured from ac susceptibility), room temperature ab-plane resistivity %%@
$\rho ^{RT}$, and zero temperature Hall coefficient $R_H ^{0}$, of LYCO films prepared with different substrate, annealing pressure p, and annealing %%@
time t.}
 \label{sample}
\begin{center}
 \begin{tabular}{c c c c c c c c c c}
\hline
sample                           & A1     & A2    & A3    & A4    & A5    & A6    & A7   &K1  \\
 \hline
 substrate                       & STO    & STO   & STO   & STO   &  STO  & STO   &STO     & KTO \\
p (10$^{-6}$Torr)                & 1      & 1     & 1     & 1     &0.8    &  0.8  & 0.6    & 1 \\
t (minutes)                      & 0      & 4     & 7     & 9     &15     &  60   & 11     & 10 \\
T$_C$ (K)                        &  0     & 0     & 4     &10    &10.5   & 10.5  & 11.2    & 5.5  \\
$\rho ^{RT} (m\Omega cm )$       &  14.2  & 10.8  & 2.0   &1.3   & -     & 3.0   & 1.9     & 1.4 \\
$R_H ^{0} (\Omega \AA /T )$      &  -220  & -100  & -19   &-11   & -     &  -    & 1       & -14  \\
\hline
 \end{tabular}
\end{center}
 \end{table}

 \begin{figure}
\includegraphics[width=7cm, height=7cm]{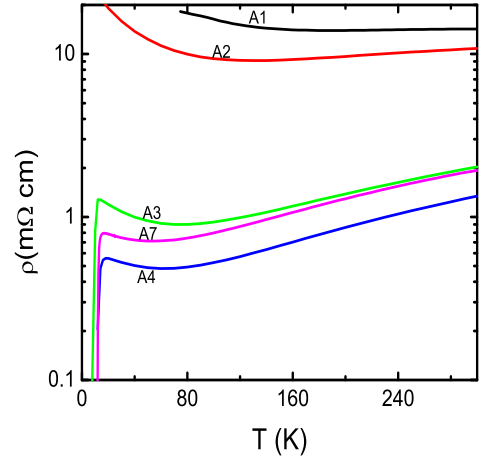}
\caption{\label{RvsT} The ab-plane resistivity of the LYCO films prepared with different annealing conditions. The growth and the annealing conditions %%@
for A1, A2, A3, A4, and A7 are shown in Table ~\ref{sample}.}
\end{figure}
 
For films grown under different oxygen conditions, the ab-plane resistivity and bulk $T_C$ are shown in Table ~\ref{sample} and in Fig.~\ref{RvsT}. %%@
The as-grown film A1, which is cooled under vacuum after deposition, shows a metallic behavior close to room temperature, followed by an %%@
insulator-like resistivity upturn at low temperatures. Longer annealing time under the same annealing pressure results in a decrease of the %%@
resistivity, by one order of magnitude, and a shift of the resistivity upturn to lower temperatures (sample A1 to A4). Superconductivity emerges and %%@
T$_C$ increases with an increasing annealing time. However, excessive annealing time ($t\ge 10$ minutes) under the same pressure does not enhance %%@
$T_C$ (see sample A5 and A6 in Table ~\ref{sample}), and severe chemical decomposition is clearly observable under an optical microscope. Indeed, %%@
better vacuum with shorter annealing time seems to further enhance $T_C$ as shown by sample A7 (Table.~\ref{sample}). Our PLD chamber can only reach %%@
$\sim 10^{-6}$ Torr, which probably limits the $T_C$ of our films, as compared with MBE-grown films \cite{Tsukada_PC_426_459}.   Comparison of $T_C$ %%@
and $\rho ^{RT}$ between  our films and the MBE-grown films \cite{Naito_SSC, Noda_PC_426_220} strongly suggests that high-vacuum annealing ($P\ll %%@
10^{-6}$ Torr) is necessary to achieve higher $T_C$'s in the LYCO system.

\begin{figure}
\includegraphics[width=7cm, height=7cm]{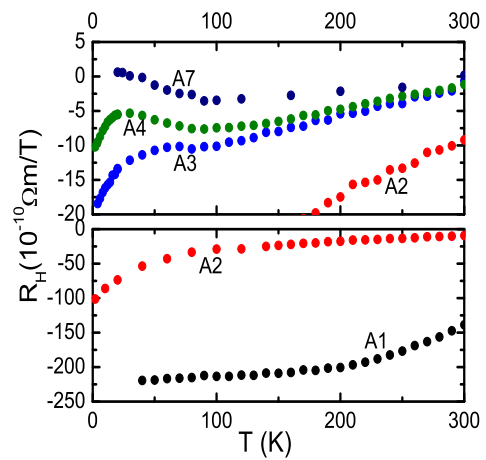}
\caption{\label{RhvsT} The Hall coefficient of the LYCO films prepared with different annealing conditions. The growth and the annealing conditions %%@
for A1, A2, A3, A4, and A7 are shown in Table ~\ref{sample}.}
\end{figure}

The normal state Hall resistivity of these films was measured with fields up to 14T. The Hall coefficients $R_H$ from room temperature down to 2K are %%@
shown in Fig.~\ref{RhvsT}. It is evident that the Hall coefficient changes dramatically with increasing annealing time. For films A1 to A3, the %%@
amplitude of $R_H$ decreases by about one order of magnitude while remaining negative, which is consistent with the decrease of resistivity shown in %%@
Fig.~\ref{RvsT} if more electron-like carriers are being created. With further annealing under higher vacuum, the Hall coefficient changes sign from %%@
negative to positive at low temperature as shown by sample A7, which suggests hole-like carriers are introduced. The overall trend of $R_H$ with %%@
annealing is similar to that of the n-doped cuprates Nd$_{2-x}$Ce$_{x}$CuO$_{4}$ (NCCO), Pr$_{2-x}$Ce$_{x}$CuO$_{4}$ (PCCO) and %%@
La$_{2-x}$Ce$_{x}$CuO$_{4}$ (LCCO) with increasing cerium doping \cite{Dagan, Sawa_prb_66_014531}. The temperature dependence of $R_H$  is also %%@
similar to these cerium-doped T'-structure cuprates.

\begin{figure}
\includegraphics[width=7cm, height=7cm]{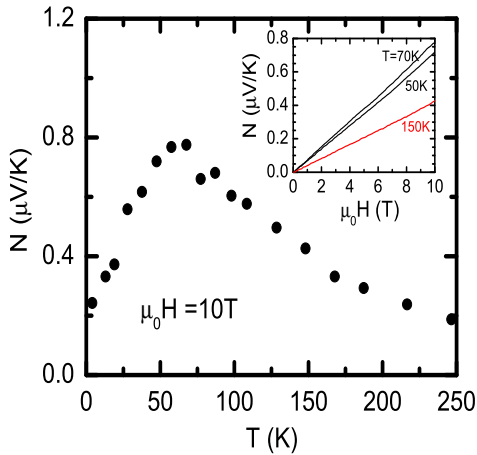}
\caption{\label{Nernst} A Nernst signal of a LYCO film (sample A5) with $T_C\approx 10.5$K. The growth and the annealing condition for A5 is shown in %%@
Table ~\ref{sample}.}
\end{figure}

To further compare with the cerium-doped T'-phase cuprates, we performed Nernst effect measurements on a 5mm$\times$10mm LYCO film (sample A5) with %%@
$T_C\approx 10.5$K. For each temperature, a constant temperature difference between two ends of the sample along the 10mm direction was stabilized by %%@
a heater. By applying a magnetic field along the c-axis, a transverse electric field $E_{y}$ was induced due to the Nernst effect. The Nernst signal %%@
N, defined as $E_{y}/ \nabla _{x}T$ where $\nabla _{x}T$ is the longitudinal temperature gradient, is shown as a function of temperature at a constant %%@
field $\mu _0 H =10$T. Above 20K, the Nernst signal shows a linear field dependence up to 10T as shown at a few temperatures in Fig.~\ref{Nernst} %%@
(inset), characteristic of a normal-state behavior. The amplitude and temperature dependence of N are comparable to results found in superconducting %%@
PCCO films \cite{Hamaza}. Below 20K, a nonlinear N with magnetic field is found (data not shown), characteristic of the superconducting state.   

\begin{figure}
\includegraphics[width=7cm, height=7cm]{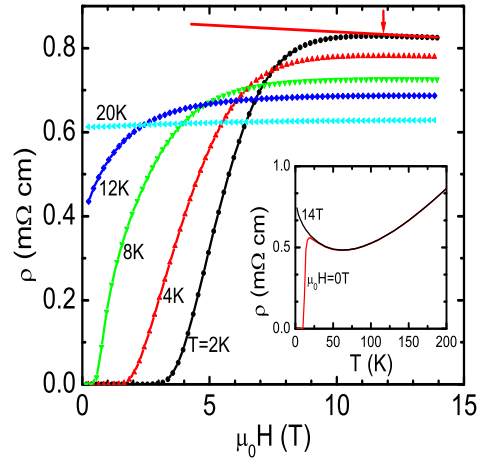}
\caption{\label{rvsh} The low-temperature transverse magnetoresistance of a LYCO film (sample A4) with $H\parallel c$. Inset: The temperature %%@
dependence of the zero-field and the normal-state ($\mu _0H=14$T) ab-plane resistivity.}
\end{figure}

Now we report the transport properties of the superconducting state. The transverse magnetoresistance of a superconducting LYCO film (sample A4) was %%@
measured at low temperatures, as shown in Fig.~\ref{rvsh}. By increasing the magnetic field to 14T, superconductivity is suppressed at all %%@
temperatures. At $T=2$K, a negative normal-state magnetoresistance emerges at high fields. The $H_{C2}$ is about 12T, which is estimated by the field %%@
where the normal-state resistivity is recovered. The negative magnetoresistance behavior at high fields and the value of $H_{C2}$ are similar to that %%@
of underdoped PCCO \cite{Fournire_prb_68_094507} and LCCO films \cite{Sawa_prb_66_014531}. At $\mu _0H=$14T, the LYCO film shows a strong resistivity %%@
upturn as the temperature decreases (Fig.~\ref{rvsh} inset), suggesting a transition from a superconductor to an ``insulator-like'' ground state, %%@
similar to underdoped superconducting LnCCO cuprates \cite{Fournire_prb_68_094507, Sawa_prb_66_014531}. 
  
These resistivity, Hall coefficient, Nernst effect, and magnetoresistance data strongly suggest that oxygen reduction in %%@
La$_{1.85}$Y$_{0.15}$CuO$_{4}$ films is equivalent to cerium doping in the Nd$_{2}$CuO$_{4}$, Pr$_{2}$CuO$_{4}$, or La$_{2}$CuO$_{4}$ systems.  The %%@
low carrier density of as-grown LYCO films, about 0.03 electrons/Cu determined by the Hall coefficient (from $R_H=1/ne$ at 2K), supports the view that %%@
LYCO originates from a doped insulator, rather than a half-filled band metal \cite{Naito_SSC, Tsukada_PC_426_459}. By oxygen reduction, both %%@
resistivity and Hall coefficient first decrease significantly, which suggests that more electron-like carriers are introduced %%@
\cite{Jiang_prb_47_8151}. By further reduction, the sign change of the Hall coefficient (sample A7) and the large Nernst signal (sample A5), suggests %%@
a Fermi surface with both electron-like and hole-like carriers, as  has been found in cerium-doped LnCCO cuprates by transport %%@
\cite{Jiang_prl_73_1291} and ARPES \cite{Armitage, Armitage2} measurements. This Fermi surface evolution from an electron-like to a two-band structure %%@
with increasing carrier density has been proposed to result from the destruction of an AFM, or SDW gap, by doping\cite{Millis, Zimmers}. The similar %%@
value of $H_{C2}$ of LYCO with that of the cerium-doped cuprates, as shown by the magnetoresistance measurements, again suggests a similar band %%@
structure and a similar mechanism for superconductivity. {\it All of our transport data suggest that LYCO originates from an antiferromagnetic %%@
Mott-like insulator and is electron-doped by oxygen reduction.} This is the main conclusion of this paper. 

In the cerium-doped cuprate superconductors, oxygen reduction has been shown to trigger the superconductivity primarily by the suppression of a %%@
disorder induced pair breaking effect \cite{Josh}. The previous work on LRCO \cite{Naito_SSC, Noda_PC_426_220, Tsukada_PC_426_459} argued for a %%@
similar disorder induced pair breaking effect and a carrier localization to explain the semiconducting resistivity of the as-grown films. The %%@
hypothesis of ref. [2-4] is that carrier doping is not necessary for superconductivity because LRCO is intrinsically a band metal. But the authors of %%@
ref. [2-4] present no quantitative evidence to support their hypothesis of carrier localization. In contrast, our transport data on LYCO suggests a %%@
significant increase of the carrier density with increasing oxygen reduction. This means that carrier doping is necessary to drive LYCO to the %%@
metallic and superconducting state. 

We now discuss oxygen reduction (by vacuum annealing) in LYCO films. It is likely that all the carriers are introduced by oxygen deficiency, because %%@
the valence of Y$^{3+}$ is not expected to change with oxygen reduction. Our best-annealed La$_{1.85}$Y$_{0.15}$CuO$_{4-\delta}$ film (sample A7) has %%@
a resistivity and Hall coefficient similar to La$_{2-x}$Ce$_{x}$CuO$_{4}$ with $x=0.08-0.10$, which would correspond to an oxygen deficiency of %%@
$\delta=0.04-0.05$ in LYCO. This amount of oxygen deficiency is not unreasonable, since an oxygen deficiency of 0.06 was found in Nd$_2$CuO$_4$ after %%@
vacuum annealing \cite{schultz}. Another point to note is that superconductivity has not been achieved in T'-structure La$_2$CuO$_4$ by oxygen %%@
reduction. Therefore, it seems that oxygen reduction is strongly affected by the rare-earth ionic environment, possibly because of the lattice %%@
distortion created by RE$^{3+}$ substitution for La$^{3+}$ \cite{Naito_SSC}. This may be similar to the enhanced oxygen intake found in %%@
La$_{2-x}$Sr$_x$CuO$_{4+\delta}$ films caused by a strain effect from the substrate \cite{Bozovic}. In this case, a significant increase in Tc was %%@
observed \cite{Bozovic, Locquet}. So it appears that the role of oxygen stoichoimetry and disorder is complicated and poorly understood in both %%@
hole-doped and electron-doped cuprates. There are issues for future study and are beyond the scope of the results we present here.
  
In summary, we have prepared superconducting T'-structure La$_{1.85}$Y$_{0.15}$CuO$_{4}$ (LYCO) films by the pulsed laser deposition for the first %%@
time. Our systematic transport studies, including resistivity, Hall effect, Nernst effect, and magnetoresistance measurements, suggest that LYCO %%@
evolves from an antiferromagnetic insulator to a two-band metal with increasing oxygen reduction. This implies that oxygen reduction in LYCO causes %%@
electron doping, and superconductivity in LYCO originates from a doped Mott-like insulator. Our results contrast with the prior proposal that (La, %%@
RE)$_{2}$CuO$_{4}$ (LRCO) is an undoped, half-filled band metal \cite{Naito_SSC, Noda_PC_426_220, Tsukada_PC_426_459}.

This work is supported by NSF under contract DMR-0352735. It is also supported by the W. M. Keck Foundation and by the NSF-UMD-MRSEC SEF. The authors %%@
would like to thank Professor A. J. Millis and Dr. J. S. Higgins for beneficial discussions.  

%\bibliography{lyco_ref}

\end{document}